\documentclass[10pt,conference,switch]{IEEEtran}
\IEEEoverridecommandlockouts
\usepackage{amsmath,amssymb,amsfonts}
\usepackage{algorithmic}
\usepackage{graphicx}
\usepackage{textcomp}
\usepackage[table]{xcolor}
\def\BibTeX{{\rm B\kern-.05em{\sc i\kern-.025em b}\kern-.08em
    T\kern-.1667em\lower.7ex\hbox{E}\kern-.125emX}}

\usepackage{enumitem}

\usepackage{todonotes}
\usepackage[numbers, sort&compress]{natbib}
\usepackage{url}
\usepackage{booktabs}
\usepackage{caption}
\usepackage{multirow}
\usepackage{censor}
\usepackage{hyperref}
\usepackage{soul}
\newcommand{\code}[1]{%
  \begingroup
  \sethlcolor{lightgray}%
  \hl{#1}%
  \endgroup
}
\usepackage{subcaption}
\usepackage{mathrsfs}
\usepackage{numprint}
\npdecimalsign{,}
\nprounddigits{2}
\usepackage{colortbl}

\newcommand{\email}[1]{\href{mailto:#1}{#1}}

\definecolor{codegreen}{rgb}{0.0, 0.5, 0.0}

\usepackage{breqn}

\begin{document}

\title{STACC: Code Comment Classification using SentenceTransformers \\
}

\author{
\IEEEauthorblockN{Ali Al-Kaswan}
\IEEEauthorblockA{\textit{Delft University of Technology} \\
Delft, The Netherlands \\
\email{a.al-kaswan@tudelft.nl}} 
\and
\IEEEauthorblockN{Maliheh Izadi}
\IEEEauthorblockA{\textit{Delft University of Technology} \\
Delft, The Netherlands \\
\email{m.izadi@tudelft.nl}}
\and
\IEEEauthorblockN{Arie van Deursen}
\IEEEauthorblockA{\textit{Delft University of Technology} \\
Delft, The Netherlands \\
\email{arie.vandeursen@tudelft.nl}}
}

\maketitle

\begin{abstract}
Code comments are a key resource for information about software artefacts. Depending on the use case, only some types of comments are useful. Thus, automatic approaches to classify these comments have been proposed.
In this work, we address this need by proposing, STACC, a set of SentenceTransformers-based binary classifiers. 
These lightweight classifiers are trained and tested on the NLBSE Code Comment Classification tool competition dataset, and surpass the baseline by a significant margin, achieving an average F1 score of 0.74 against the baseline of 0.31, which is an improvement of 139\%.
A replication package, as well as the models themselves, are publicly available.
\end{abstract}


\section{Introduction}
\label{introduction}
Code comments are an essential source of information for software developers. Code comments document functionalities, describe the usage, or are a place for developers to leave development notes, which help with maintainability~\cite{kostic2022code} and ongoing development tasks~\cite{DiSorboVPCP21}. Code comments also play a large role in helping developers understand source code~\cite{kostic2022code}. 

Depending on the task at hand, not all types of code comments are useful. Especially when considering automated code tasks, like automatic code summarization~\cite{al2023extending}, or code completion~\cite{izadi2022codefill}, only certain types of documentation are useful. Therefore, there is a need to classify to which category a certain piece of software documentation belongs. 

In this tool paper, we present \textbf{STACC}: a set of \textbf{S}entence\textbf{T}ransformer-\textbf{A}ssisted~\cite{reimers-2019-sentence-bert} \textbf{C}omment \textbf{C}lassifiers. We make use of SetFit~\cite{setfit}, an efficient and prompt-free framework for few-shot fine-tuning of Sentence Transformers. SetFit achieves high accuracy with little labelled data. The resulting models are also compact and fast to run.

STACC outperforms the baseline in every category in both recalls, as well as F1 scores. We achieve an average F1 score of 0.74 against the baseline of 0.31.

We make all of our models available on the HuggingFace Hub.\footnote{HF Hub:~\url{https://huggingface.co/AISE-TUDelft}} The training Notebooks, which explain the model selection and training process in detail and show how to load and use STACC are available on GitHub and Google Colab\footnote{STACC Repo:~\url{https://github.com/AISE-TUDelft/STACC}}. Finally, we also offer an interactive HuggingFace Space to experiment with our classifiers in the browser.\footnote{STACC HF Space:~\url{https://huggingface.co/spaces/AISE-TUDelft/STACC}}

\section{Tool Construction}
In this section, we provide details on how we constructed STACC. Specifically, how we selected the base model, and how we tuned the hyperparameters to create the set of classifiers.

\subsection{Model Selection}
We first test several candidate base models for our approach. We select the models from the Sentence-Transformers Sentence Embedding benchmark.\footnote{Sentence-Transformers Model Overview: \url{https://www.sbert.net/docs/pretrained_models.html\#model-overview}} 

We tune the models on a sample of 32 positive and negative examples from a single category, for 5 epochs using the standard settings in the SetFitTrainer. We select the hardest category, the one where the baseline performed poorest, to tune our hyperparameters. We measure the accuracy and F1 on the complete test set as well as the total runtime of the training and testing regime. 

We first select two smaller models.~\code{paraphrase-MiniLM-L3-v2} which is specifically designed for semantic tasks and has a model size of only 61MB.~\code{all-MiniLM-L6-v2}, a good all-around model, with a model size of 120MB. 
We also select~\code{all-mpnet-base-v2}, which is the highest-performing model in the Sentence Embedding benchmark but is much larger with a model size of 420MB. 
Finally, we also selected~\code{st-codesearch-distilroberta-base}, in the hope that its mixed training objectives, which included some code and documentation-related tasks would boost its performance. Its model size was 320MB.
We attempt to run~\code{all-roberta-large-v1} but we find that this model was too big to run on our hardware with a model size of 1.4GB.  

\subsection{Hyperparameter Tuning}
We tune the hyperparameters of the selected model trained on one category. We make use of the Optuna backend with SetFit to find the best hyperparameters. We tune the learning rate, the number of training epochs, the number of sentence-pairs generation iterations and the solver. We use make use of the entire dataset and run Optuna for 20 iterations to find the best hyperparameters. Finally, we use the selected base model and optimised hyperparameters to train a classifier for each of the 19 categories. 

\section{Experimental Setup}
\label{experimental}
In this section, we describe the dataset, metrics and baseline provided by the competition, as well as the implementation details of STACC.

\subsection{Dataset}
We utilise the dataset provided by the NLBSE Code Comment Classification competition.\footnote{Code Comment Classification GitHub repo: \url{https://github.com/nlbse2023/code-comment-classification}} This dataset consists of 3 languages, Java, Python and Pharo. For each language, the authors selected several projects and extracted the documentation. The authors have taken the documentation, split it into sentences, and manually labelled each sentence. Asides from the sentence, the dataset also denotes from which file the comment was extracted. We simply append this to the comment sentence and separate the two using a `pipe` symbol. 

For example: \code{Method to calculate the SHA-256 checksum  
\textbar Checksum.java} 

Each language has several categories to which a sentence can belong. Java, Python and Pharo have 7, 5 and 7 categories respectively~\cite{rani2021}.

\begin{table}[tb]
    \centering
    \begin{tabular}{ll|ll|ll|l}
        \noalign{\smallskip}\toprule
        Language & Category & Train & ~ & Test & ~ & Total \\ 
        \midrule
        ~ & ~ & Pos & Neg & Pos & Neg & ~ \\ 
        \toprule
        Java & Expand       & 505 & 1426 & 127 & 360 & 2418 \\
        Java & Ownership    & \textbf{90} & 1839 & 25 & 464 & 2418 \\ 
        Java & Deprecation  & \textbf{100} & 1831 & 27 & 460 & 2418 \\ 
        Java & Rational     & 223 & 1707 & 57 & 431 & 2418 \\ 
        Java & Summary      & 328 & 1600 & 87 & 403 & 2418 \\ 
        Java & Pointer      & 289 & 1640 & 75 & 414 & 2418 \\ 
        Java & Usage        & 728 & 1203 & 184 & 303 & 2418 \\
        \midrule
        Pharo & Resp        & 267 & 1139 & 69 & 290 & 1765 \\ 
        Pharo & Keymsg      & 242 & 1165 & 63 & 295 & 1765 \\ 
        Pharo & Keyimpl     & 184 & 1222 & 48 & 311 & 1765 \\ 
        Pharo & Collaborators & \textbf{99} & 1307 & 28 & 331 & 1765 \\ 
        Pharo & Example     & 596 & 812 & 152 & 205 & 1765 \\ 
        Pharo & Classref    & \textbf{60} & 1348 & 17 & 340 & 1765 \\ 
        Pharo & Intent      & 173 & 1236 & 45 & 311 & 1765 \\ 
        \midrule
        Python & Expand     & 402 & 1637 & 102 & 414 & 2555 \\
        Python & Parameters & 633 & 1404 & 161 & 357 & 2555 \\
        Python & Summary    & 361 & 1678 & 93 & 423 & 2555 \\ 
        Python & Devnotes   & 247 & 1792 & 65 & 451 & 2555 \\ 
        Python & Usage      & 637 & 1401 & 163 & 354 & 2555 \\
        \bottomrule
    \end{tabular}
    \caption{Dataset Properties}
    \label{tab:data}
\end{table}

The dataset consists of samples, each sample has an id, the sentence text, the class or file from which the sentence was extracted, a partition (train or test), a category, and the instance type (1 for a positive instance, 0 otherwise).

The dataset is quite unbalanced for each category, the negative samples far outnumber the positive ones, as can be seen, in~\autoref{tab:data}. We have identified the particularly unbalanced classes and marked them in \textbf{bold}, these will be revisited in~\autoref{results}.

There are around 2.4K training samples for each Java and 2.6K for each Python category, while there are only about 1.8K for each Pharo category. This limits us in the type of models used for this task since data-hungry models cannot be trained with this amount of data.

\subsection{Metrics}
To score the classifiers we use the metrics specified by the competition as a function of a category \(c \in C\). Namely, the recall \(R_c\), precision \(P_c\), and F1 score \(F_{1,c}\). These scores are calculated per category \(c\). The F1 score is calculated as the geometric mean of the recall and precision, we additionally use the weighted F1 score, \(wF_{1,c}\) as well as the accuracy to also take the unbalanced nature of the data into account, but these scores are just used as a reference and are not taken into account for the final scoring.
The metrics are defined as follows:
\begin{equation}
    \begin{aligned}
    Accuracy_c &= \frac{TP_c + TN_c }{TP_c + TN_c + FP_c + FN_c} \\
    P_c &= \frac{ TP_c }{ TP_c + FP_c } \\ 
    R_c &= \frac{ TP_c }{ TP_c + FN_c } \\
    F_{1,c} &= 2 \cdot \frac{ P_c \cdot R_c }{ P_c + R_c }
    \label{eq:f1}
    \end{aligned}
\end{equation}

Finally, the competition submission score for \(M\) a set of models \(m_c\) against the baseline  is defined as:
\begin{equation}
    \begin{aligned}
    submission\_score(M) &= \frac{\sum_{c \in C} F_{1,c}}{|C|} \times 0.75 \\
    &+ \newline (\%\_outperformed\_categories) \\
    &\times 0.25
    \label{eq:subscore}
    \end{aligned}
\end{equation}
This score takes the average F1-score achieved by the classifiers and weighs it against the fraction of classifiers that outperform the baselines.

\subsection{Baselines}
The authors of the competition also provide baseline Random Forest Classifiers~\cite{nlbse2023}. We observe quite poor recalls but decent wF1 scores across the board in \autoref{tab:res}. This indicates that the model tends to favour predicting non-membership for each of the classes, which the unbalanced nature of the data could explain. The category with the lowest F1 score is `Java Depreciation`, this is the category which we target for the model selection and Hyperparameter search. 

\subsection{Implementation}
We use the HuggingFace SentenceTransformers and PyTorch packages to train STACC. The base model was loaded using the HuggingFace Hub. To process the data we use Pandas and the HuggingFace Dataset packages. The experiments are conducted on an Nvidia GTX 3080 GPU with 10GB of VRAM and an AMD Ryzen Threadripper 3990X paired with 128GB RAM. 

\section{Results}
\label{results}
In this section, we discuss the results of our model search and hyperparameter tuning. We also compare our models to the baseline and provide a short discussion of the results.

\subsection{Model Selection}
\begin{table}[tb]
    \centering
    \begin{tabular}{l|lll}
        \noalign{\smallskip}\toprule
        Model Name                          &   Accuracy    &   F1  &  Time  \\
        \toprule
        paraphrase-MiniLM-L3-v2             &   0.88        &  0.435 &   1:22 \\
        all-MiniLM-L6-v2                    &   0.89        &  0.453 &   2:30 \\
        all-mpnet-base-v2                   &   0.93        &  \textbf{0.646} &   4:37 \\
        st-codesearch-distilroberta-base    &   0.88        &  0.457 &   2:41 \\
    \end{tabular}
    \caption{Accuracy, unbalanced F1 score and training time in minutes of different base models on the Java Depreciation category}
    \label{tab:models}
\end{table}

The results of the model selection experiments are shown in \autoref{tab:models}, and the training time is denoted in minutes. We can see that \code{all-mpnet-base-v2} outperforms the other models by a decent margin. Its training time is longer but still workable. Unfortunately, the \code{st-codesearch-distilroberta-base} model did not seem to benefit from its pre-training regime despite being slower and larger than both \code{MiniLM} models.

\subsection{Hyperparameter Tuning}
\begin{figure}[tb]
    \centering
    \includegraphics[width=\linewidth]{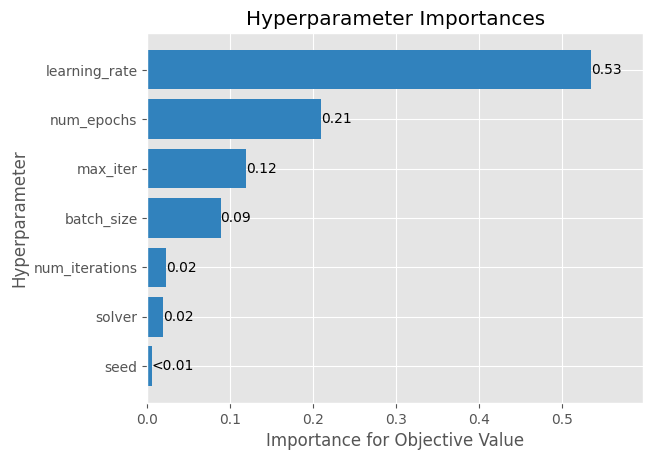}
    \caption{fANOVA importance of each hyperparameter}
    \label{fig:hyper}
\end{figure}

The importance of each hyperparameter is plotted in \autoref{fig:hyper}. The importance of each hyperparameter is calculated using the fANOVA importance metric included in the Optuna backend of SetFit~\cite{hutter2014efficient}. The obtained hyperparameters, the space which we searched and the base settings are shown in ~\autoref{tab:hyper}. We find that the most important hyperparameter is the learning rate, followed by the number of training epochs. 
  
\begin{table}[tb]
    \centering
    \begin{tabular}{l|ll|l}
        \noalign{\smallskip}\toprule
        Parameter               & Search space &   Base    & Ours          \\
        \toprule
        Learning rate           &  (1e-6, 1e-4) &   2.00e-5 & 1.71e-05      \\
        Epochs                  &  (1, 10)      &     5     & 6             \\
        LogReg fit iterations   &  (50, 300)    &     100     & 241            \\
        LogReg Solver           &  [newton-cg, lbfgs, liblin] &  liblin   & lbfgs         \\
        \bottomrule
    \end{tabular}
    \caption{Hyperparameter search space and results}
    \label{tab:hyper}
\end{table}

\subsection{STACC Models}
Finally, we use the selected base model and optimised hyperparameters to train a classifier for each of the 19 categories. Training each classifier with the full training data takes around 3 hours.
\begin{table*}[tb]
    \centering
    \begin{tabular}{ll|cc>{\columncolor[gray]{0.8}}cc|cc>{\columncolor[gray]{0.8}}cc|>{\bfseries}c}
        \noalign{\smallskip}\toprule
        ~ & ~ & \multicolumn{3}{l}{Baseline}  & ~  & \multicolumn{3}{l}{STACC}  & ~ & ~ \\
        \toprule
        Language & Category & \(P_c\) & \(R_c\) & \(F_{1,c}\) & \(wF_{1,c}\) & \(P_c\) & \(R_c\) & \(F_{1,c}\) & \(wF_{1,c}\) & \(\Delta F_{1,c}\) \\
        \toprule
        Java & Deprecation  & 0.00 & 0.00 & 0.00 & 0.92 & 0.78 & 0.95 & 0.86 & 0.98 & +0.86  \\ 
        Java & Expand       & 0.35 & 0.27 & 0.30 & 0.66 & 0.71 & 0.80 & 0.75 & 0.88 & +0.45  \\
        Java & Ownership    & 1.00 & 0.68 & 0.81 & 0.98 & 1.00 & 1.00 & 1.00 & 1.00 & +0.19  \\ 
        Java & Pointer      & 0.67 & 0.24 & 0.35 & 0.84 & 0.71 & 0.82 & 0.76 & 0.93 & +0.40  \\ 
        Java & Rational     & 0.63 & 0.30 & 0.40 & 0.88 & 0.81 & 0.92 & 0.86 & 0.97 & +0.46  \\ 
        Java & Summary      & 0.38 & 0.29 & 0.33 & 0.78 & 0.85 & 0.76 & 0.80 & 0.93 & +0.48  \\ 
        Java & Usage        & 0.54 & 0.36 & 0.43 & 0.62 & 0.83 & 0.89 & 0.86 & 0.90 & +0.43  \\ 
        \midrule
        Pharo & Classref    & 0.33 & 0.06 & 0.10 & 0.93 & 0.47 & 0.57 & 0.52 & 0.96 & +0.42  \\ 
        Pharo & Collab      & 0.47 & 0.25 & 0.33 & 0.91 & 0.36 & 0.91 & 0.51 & 0.94 & +0.19  \\ 
        Pharo & Example     & 0.77 & 0.43 & 0.55 & 0.68 & 0.93 & 0.89 & 0.91 & 0.92 & +0.35  \\ 
        Pharo & Intent      & 0.58 & 0.33 & 0.42 & 0.87 & 0.87 & 0.89 & 0.88 & 0.97 & +0.45  \\ 
        Pharo & Keyimpl     & 0.18 & 0.10 & 0.13 & 0.79 & 0.69 & 0.79 & 0.73 & 0.93 & +0.60  \\ 
        Pharo & Keymsg      & 0.31 & 0.16 & 0.21 & 0.76 & 0.79 & 0.91 & 0.85 & 0.95 & +0.64  \\ 
        Pharo & Resp        & 0.59 & 0.33 & 0.43 & 0.81 & 0.67 & 0.63 & 0.65 & 0.86 & +0.22  \\ 
        \midrule
        Python & Devnotes   & 0.17 & 0.17 & 0.17 & 0.79 & 0.43 & 0.54 & 0.48 & 0.88 & +0.31  \\ 
        Python & Expand     & 0.26 & 0.20 & 0.22 & 0.72 & 0.52 & 0.56 & 0.54 & 0.82 & +0.31  \\ 
        Python & Parameters & 0.51 & 0.22 & 0.31 & 0.65 & 0.78 & 0.86 & 0.81 & 0.89 & +0.50  \\ 
        Python & Summary    & 0.12 & 0.08 & 0.09 & 0.71 & 0.62 & 0.64 & 0.63 & 0.87 & +0.54  \\ 
        Python & Usage      & 0.47 & 0.18 & 0.26 & 0.63 & 0.69 & 0.77 & 0.73 & 0.84 & +0.46  \\ 
        \toprule
        Average & ~         & 0.44 & 0.24 & 0.31 & 0.79 & 0.71 & 0.79 & 0.74 & 0.92 & +0.43  \\ 
        \bottomrule
    \end{tabular}
    \caption{Performance of STACC against the baseline}
    \label{tab:res}
\end{table*}

We plot the results in \autoref{tab:res}. We find that the average F1 score of STACC is much higher than the baseline classifiers, with scores of 0.74 and 0.31 respectively. Each classifier has a higher F1 score and recall than its baseline counterpart.

STACC's moderately higher precision (0.44 vs 0.71) and considerably higher recall (0.24 vs 0.79) indicate that STACC tends to make more predictions for the member class. 

\vspace{2em}

With these results, per \autoref{eq:subscore} we achieve a final submission score of:
\begin{align}
submission\_score = 0.74 * 0.75 + 1 * 0.25 = 0.81
\end{align}

\subsection{Discussion}
When comparing languages it becomes apparent that STACC tends to perform better on Java than both Python and Pharo. Across the languages, the average F1 scores are 0.84, 0.72 and 0.64.

For some categories, we see that our approach greatly outperforms the baseline. For instance,~\code{Java Depreciation} and~\code{Python Summary}, where the average F1 score is 0.86 and 0.54 higher, respectively.

While for others we observe that both our and the baseline classifiers perform quite poorly. Namely,~\code{Python Devnotes},~\code{Pharo Classreferences} and~\code{Pharo Collaborators}, which are very unbalanced and have relatively few positive instances. Another imbalanced category with few positive instances is ~\code{Java Ownership}, but our classifier performed admirably in this case, achieving a perfect F1 score. Looking at the data, we notice that each of the~\code{Java Ownership} comments contains the token~\code{@author}, which makes predicting these quite straightforward. 

We also observed that the addition of the classname to the input of the model improves the average precision, recall, and F1 scores by 0.12, 0.02, and 0.09 respectively. The intuition is that the classname is in itself an extremely brief summary of the entire class. For brevity, we have omitted the results without the classnames, but the fully trained Classifiers are also available on the HuggingFace Hub. Each classifier has two branches: `main` for the regular model which expects inputs including the classname, and a branch tagged `V1` for inputs without the classname.\footnote{STACC V1 example: \url{https://huggingface.co/AISE-TUDelft/java-rational-classifier/tree/V1}}

While we focus on creating the best-performing classifiers for this task, a significant speedup can be achieved by reducing the number of training epochs. Instead of training for 6 epochs, tuning the models for 1 or 2 epochs will perform slightly worse than STACC's classifiers while only requiring a fraction of the training time.

Besides just reducing the number of training epochs, the models could also be fine-tuned in a few-shot setting by taking a few positive and negative classes. In our evaluations, we however find that a reduction in the number of training samples leads to a very large performance penalty. So in general, we recommend training with more samples but fewer iterations to reduce the training time most effectively.  

Finally, the hyperparameter search was only performed on a single category with a sample of the training data. This might bias the results. A hyperparameter search with the full training data would be preferable, but computationally quite expensive.

\section{Conclusion}
\label{conclusion}
We present, STACC, a set of binary classifiers built on the SetFit~\cite{setfit} framework. Our solution is trained and tested on the NLBSE Code Comment Classification competition and greatly outperforms the average baselines in average recall, precision and F1 scores. Furthermore, STACC improves over the baseline classifiers in recall and F1 for each category. These results show that Sentence-Transformers is an efficient and performant architecture to train classifiers for code-related tasks. All of the code used to construct STACC, as well as the models themselves are publicly available. 

\bibliographystyle{IEEEtranN}
\bibliography{references.bib}

\end{document}